# Threshold fields for antiparallel ferroelectric domain wall motion

Samrat Choudhury[1], Yulan Li[1], Nozomi Odagawa[2], Aravind Vasudevarao[1], L. Tian[1], Pavel Capek[3], Volkmar Dierolf[3], Anna N. Morozovska[4], Eugene A. Eliseev[5], Long-qing Chen[1], Yasuo Cho[2], Sergei Kalinin[6], and Venkatraman Gopalan[1,*]

[1] Materials Science and Engineering, Pennsylvania State University, University Park, PA 16802

[2] Research Institute of Electrical Communication, Tohoku University, 2-1-1 Katahira Aoba-ku Sendai 980-8577, Japan

[3] Lehigh University, Physics Department, 16 Memorial Drive, Bethlehem, PA 18015, USA

[4] V. Lashkarev Institute of Semiconductor Physics, National Academy of Science of Ukraine, 41, pr. Nauki, 03028 Kiev, Ukraine

[5] Institute for Problems of Materials Science, National Academy of Science of Ukraine, 3, Krjijanovskogo, 03142 Kiev, Ukraine

[6] Materials Science and Technology Division and Center for Nanophase Materials Sciences, Oak Ridge National Laboratory, Oak Ridge, TN 37831

---

[*] email: vgopalan@psu.edu

While an ideal antiparallel ferroelectric wall is considered a unit cell in width (~0.5nm), we show using phase field modeling that the threshold field for moving this wall dramatically drops by 2-3 orders of magnitude if the wall were diffuse by only ~2-3nm. Since antiparallel domain walls are symmetry allowed in all ferroelectrics, and since domain wall broadening on nanometer scale is widely reported in literature, this mechanism is generally applicable to all ferroelectrics.

Topological defects in materials play a critical role in understanding their real-world physical behavior. For example, dislocations explain the low deformation stresses required to overcome the otherwise large intrinsic Peierls potential barrier predicted for a perfect lattice[1,2]. Additionally, the threshold stress to move a dislocation is inversely proportional to the spatial extent of the local stress field around a dislocation[3]. Similarly, the coercive field to move a magnetic domain wall decreases exponentially as the wall width increases[4]. Thus, one might expect similar trends in ferroelectrics, which contain two or more switchable states of built-in electrical polarization under the application of an electric field. However unlike magnetic walls, a ferroelectric domain wall is theoretically predicted to possess an intrinsic width on a unit-cell level (~0.5nm).[5,6] The classical ferroelectrics literature thus predominantly treats the wall as an infinitely sharp plane with no physical extent, and the domain-reversal process as primarily a nucleation driven problem at "defects". An important outstanding issue in domain reversal in ferroelectrics is the orders of magnitude difference between the theoretically predicted coercive fields, $E_c$, of >1000kV/cm,[7] and the experimental fields that are typically on the order of 1-10 kV/cm. Attempts to explain this discrepancy has been based on nucleation assisted domain growth models.[8,9,10] Miller and Weinreich theory[11] proposes the lateral motion of an atomically sharp domain wall through preferential nucleation of a domain nucleus at the wall. However, these models ignore the internal structure of the wall, and thus predict a zero threshold coercive field when $t\to\infty$ (bulk crystals), which is experimentally incorrect.[9] In order to account for the experiments, these models postulate (without modeling), a *threshold field*, $E_h$, to move a domain wall.

Suzuki and Ishibashi[12] and later Sidorkin[13] have shown that the threshold field for a ferroelectric domain wall arising from the Peierls barrier of a 1-D lattice decreases exponentially as a function of increasing domain wall width. Recent works by Catalan et. al.[14] and Shin et. al.[15] have explored the role of internal structure of the domain wall itself on the domain reversal process. Previously, Bandyopadhyay and Ray[16] predicted the upper limit for $E_h$ for domain walls of finite width, $2\omega_o$ as $E_h \leq a\alpha_{33}P_s/\omega_o$. For the uniaxial trigonal (*3m*) ferroelectric lithium niobate (LiNbO$_3$), which is the focus of this work, the Landau coefficient $\alpha_{33} = 1/2\varepsilon_{33} \sim 2\times10^9 \text{Nm}^2/\text{C}^2$, where $\varepsilon_{33}$ is the dielectric permittivity, the spontaneous polarization $P_s$~0.75 C/m$^2$, and the lattice parameter, $a$ =0.515nm, by which a 180° wall moves laterally. Thus, according to [9], $E_h \leq 30000$ kV/cm for a unit cell sharp domain wall, $2\omega_o=a$. In contrast, the experimental coercive fields for LiNbO$_3$ are typically in the range of $E_c$~ 2 kV/cm (stoichiometric composition) to 210 kV/cm (for congruent composition). However, even in non-stoichiometric (congruent) LiNbO$_3$ where $E_c$=210 kV/cm, the threshold field to move domain walls pinned between defect sites has been observed to be $E_h \leq 15$ kV/cm, though it is only an upper limit.[17] It is thus important to distinguish the average coercive field $E_c$, which can be influenced by domain wall pinning events, and the threshold coercive field $E_h$ to locally move a wall, which can be much smaller than $E_c$ and is the subject of this study. The typical range for $E_h$ in all compositions of LiNbO$_3$ and LiTaO$_3$ is ~0.5-15 kV/cm. In this work, we numerically simulate the relationship between $E_h$ and wall width, $2\omega_o$, for two specific materials, LiNbO$_3$ and LiTaO$_3$. We show that even a broadening to $2\omega_o$~2-3nm can exponentially lower the $E_h$ in these materials; with a second exponential dependence for wall widths than are even wider.

The complete analytical Ginzburg-Landau-Devonshire (GLD) total free energy for the prototype paraelectric phase ($\bar{3}m$) of LiNbO$_3$ and LiTaO$_3$ in terms of order polarization vector $P_i$ and strain tensor $\varepsilon_{ij}$ is given by,

$$F = \int (-\frac{\alpha_{ij}P_iP_j}{2} + \frac{\beta_{ijkl}P_iP_jP_kP_l}{4} + \frac{C_{ijkl}\varepsilon_{ij}\varepsilon_{kl}}{2} - \gamma_{ijkl}\varepsilon_{ij}P_kP_l + \frac{g_{ij}}{2}\left(\frac{\partial P_j}{\partial x_i}\right)^2 - \frac{P_iE_{i,dd}}{2} - P_iE_{i,ext})dV$$
(1)

where, $\alpha_{ij}$ and $\alpha_{ijkl}$ are the first and second order impermittivity tensors, $C_{ijkl}$, $\gamma_{ijkl}$, and $g_{ij}$ are the elasticity, electrostrictive, and gradient tensors, respectively, The numerical values for all these quantities in LiNbO$_3$ and LiTaO$_3$ are given in Ref [18]. Further, $\varepsilon_{ij}$ is strain, $E_{i,ext}$ is the external electric field, $E_{i,dd}$ the dipole-dipole interactions field, and $V$ is the simulation volume. A single infinite domain wall with a polarization profile $P = P_s \tanh(x/\omega_o)$ is defined, where $P_s$ is the spontaneous saturation polarization, $x$ is the coordinate normal to the wall, and coordinate $z$ is parallel to $P_s$. The temporal and spatial evolution of the polarization vector field was obtained by minimizing the energy in Eq (1) by solving the time-dependent GLD equations using the semi-implicit Fourier spectral method. A single 180° domain wall was placed in the simulation volume of $128a \times 2a \times 128a$ for a smaller gradient coefficient or $512a \times 2a \times 128a$ for a large gradient coefficient when the wall width $2\omega_o$ increased above ~2nm. The film thickness was taken as $t=na$ with $n<128$. Periodic boundary conditions were employed in $x_1$ and $x_2$ directions. Two cases were considered: **Case 1**: a single domain wall in an infinite ferroelectric medium with no surfaces, and **Case 2**: a single wall in a ferroelectric of finite film thickness, $t$, in the polarization direction with top and bottom electrodes. The threshold field was determined as the field $E_h$ needed to move the wall by one unit cell distance, $a$, laterally. The lattice friction felt by a moving wall is approximated by discretizing the continuum GLD equation using a grid size that is equal to the lattice spacing, $a$, in the crystal. Since we find the wall width, $\omega_o$ to have experimental variability in this work, and since $g_{ij}$ is typically determined from experimental wall width, as $\omega_o \sim \sqrt{2g_{13}/\alpha_{33}}$, we vary *only* the gradient coefficient $g_{13}$ in the simulation, while keeping *all* other material properties the same. (This point is discussed again further on).

Figure 1(a) shows the threshold field, $E_h$ versus wall width, $2\omega_o$ for the two cases. A striking observation is that even a small broadening of $2\omega_o$~2-3nm can dramatically lower coercive fields in these materials for both Cases 1 and 2.[19] Note that while the wall width is uniform for Case I, the wall broadens at the surface for Case 2 as shown in the inset, where $t=96a$ was assumed. However, the $E_h$ is in excellent agreement between Cases 1 and 2, when the wall width $2\omega_o$ at $z=t/2$ is plotted for Case 2 in Figure 1(a). Thus an important conclusion is that for thick films, the bulk wall width determines the $E_h$, suggesting that the bulk of the wall exerts a drag on the surface triple junction under an external field. A better correlation is between $E_h$ and the integrated average wall width, given by $2\int \omega_o(z)dz/t$, and plotted in Figure 1(b) for a fixed gradient coefficient $g_{13}$. Only in very thin films does the surface broadening of the wall influence the $E_h$ as seen in Figure 1(b). Note that while conventional nucleation models predict an *increase* in the coercive field, $E_c$ with thinner films, the threshold field $E_h$ is predicted to remain unchanged for thicknesses much greater than the surface wall broadening depth, *d*, and *decreases* only when *t* approaches *d*. Since in congruent LiTaO$_3$ sample, the coercive field $E_c$ ~210kV/cm *does not* change over a thickness range of 500nm to 0.5mm,[20] we expect that the nucleation model in Ref [9] is not the dominant mechanism for explaining the observed coercive field, $E_c$ in these materials. Instead, the growth of preexisting domain walls in the crystals requiring a threshold field $E_h$ to grow and a depinning field $E_c$ to overcome pinning sites appears more relevant. This would also predict that the threshold coercive field will decrease for thinner single crystal LiNbO$_3$ and LiTaO$_3$ of *t*~1-10nm due to surface effects.

Figure 1(a) predicts that experimental threshold fields $E_h$ can be explained if domain walls were of the order of ~1.5-2nm (congruent composition) to ~10nm (stoichiometric composition) in LiNbO$_3$ and LiTaO$_3$. Is it reasonable? Broadening of a domain or twin wall on nanometer scale due to extrinsic defects[21,22,23] charged walls,[24] and surfaces[25,26] has been observed in other material systems before. Direct imaging of strain,[27,28] index contrast[29], and other optical properties[30,31] at domain walls reveal property changes on length scales of 1-30μm.[32] On the nanoscale, scanning nonlinear dielectric microscopy (SNDM) of lithium tantalate,[33] performed in cross-sectional, *y*-cut geometry of the stoichiometric (congruent) crystal reveals that the wall width is of the order of 5.5nm at ~5nm depth and decreases inside the crystal to a width of ~2.5nm at a depth of ~100 nm from the $z$-surface of the crystal. Daimon and Cho[33] also suggest that the "average" wall width in stoichiometric $LiTaO_3$ is about 3 times narrower than in congruent composition. This may appear to be in contradiction with the higher coercive fields $E_c$ for the congruent composition. However, as argued before, we believe that the threshold field $E_h$ itself is similar for these two compositions, while the large difference in the coercive field $E_c$ between the compositions comes from the well-established increased wall-pinning events in congruent composition due to increased pinning defects. Also the actual wall width in both compositions shows a wide range of values spatially, from 20-100nm. The surface broadening of the wall over 20-100nm in lithium niobate is also supported by a combined experimental and theory investigation of these walls using piezoelectric force microscopy.[34] Interestingly, SNDM shows that wall broadening at ferroelectric surfaces can be quite significant (factor of ~10-100 times the bulk wall width) in real crystals; certainly larger than what phase field modeling predicts for electrically neutral surfaces (factor of ~2 times the bulk wall width). This appears to be a sensitive function of the nature of the surfaces, including its electrostatic boundary condition, preparation, and defects; this is the subject of a separate study.[35]

By varying the gradient coefficient, $g_{13}$, the domain wall energy $\sigma \propto \sqrt{g_{13}\Delta f}$ also changes, where $\Delta f \sim \alpha_{33}^2/4\beta_{3333}$ is the Landau energy barrier. In order to determine if the change in threshold coercive field, $E_h$ is due to a change in the wall width, $2\omega_o$ or the wall energy, $\sigma$, we repeated the simulation in Figure 1(a) by performing the following two numerical tests: (1) Fix $\sigma$ and $P_s$, while varying wall width $2\omega_o$: By arbitrarily scaling $g_{13}$ by a constant $K$ ($K>1$), while scaling $\alpha_{33}$, and $\beta_{3333}$, by $1/K$, we linearly scale the wall width, $\omega_o \sim P_s\sqrt{g_{13}/\Delta f}$ by a factor $K$, while keeping the wall energy $\sigma$ and the saturation polarization $P_s = \sqrt{\alpha_{33}/\beta_{3333}}$ constant. The results clearly show a dramatic drop in $E_h$ with scaling of the wall width in a bulk crystal. (2) Fix $2\omega_o$ and $P_s$, while varying wall energy $\sigma$. If $g_{13}$, $\alpha_{33}$, and $\beta_{3333}$ are all scaled by $K$, the wall width $\sigma$ and polarization $P_s$ remain the same, while the wall energy scales linearly by $K$. This leads to a linear *increase* in the threshold coercive field for a single domain wall in a bulk crystal. Though scaling $\alpha_{33}$ and $\beta_{3333}$ in the above numerical tests changes the material itself, these tests do confirm the central role of wall width instead of wall energy in the decrease of threshold coercive field in Figure 1(a).

In conclusion, we show a dramatic decrease of 2-3 orders of magnitude in $E_h$ for a wall diffuseness of even 2-3nm, which would bring it in closer agreement with the experimental threshold fields for wall motion in lithium niobate and tantalate. A switching mechanism through preexisting, slightly diffuse antiparallel ferroelectric walls can be general to all ferroelectrics, and can play a significant role in determining the experimental coercive fields in ferroelectrics.

**Acknowledgments**: We would like to gratefully acknowledge NSF-grant numbers, DMR-0507146, DMR-0512165, DMR-0349632, DMR-0213623, DMR-0602986, DMR-0507146, ARO-Grant W911NF-04-1-0323 and DOE grant number DE-FG02-07ER46417. Research was also sponsored in part by the Division of Materials Sciences and Engineering, Office of Basic Energy Sciences, U.S. Department of Energy, under

**Figure Captions:**

**Figure 1: (a) Phase field modeling of the threshold coercive field, $E_h$ for the motion of a single domain wall in LiNbO$_3$ (LN) and LiTaO$_3$ (LT) versus domain wall width, $2\omega_o$. The film was $t=96a$ thick, and the simulation size was $128a \times 2a \times 128a$ for $2\omega_o$ <2nm and $512a \times 2a \times 128a$ for $2\omega_o$ >2nm, where $a=0.515$nm. Inset shows the polarization profile at the junction between the wall and one of the surfaces of the film. The bulk simulation was infinite in all dimensions. (b) The dependence of $E_h$ on film thickness for a fixed $g_{13}=4\alpha_{33}a^2$. Also the domain wall, $2\omega_o$ at $z=0$, $z=t/2$ and an average quantity, $2\int \omega_o(z)dz/t$ are plotted.**

**Figure 2: Scanning nonlinear dielectric microscopy images of a circular domain in a 40nm thick z-cut single crystal lithium niobate (a,b) and a 31nm thick z-cut single crystal lithium tantalate, (c,d) at first harmonic, $\omega_p$ =6kHz (a,c), and second harmonic, $2\omega_p$ =12kHz (b,d) modulation frequencies. Figure 2d, e show typical line profiles from these images as labeled, and pairs of arrows indicate the wall region. Images (a-d) are plotted in 3-dimensional orthographic view with a ~21° rotation about the horizontal axes of the images. The length scales directly correspond to the horizontal axes of the images. The same domain region is imaged in (a) and (b), and similarly in (c) and (d).**

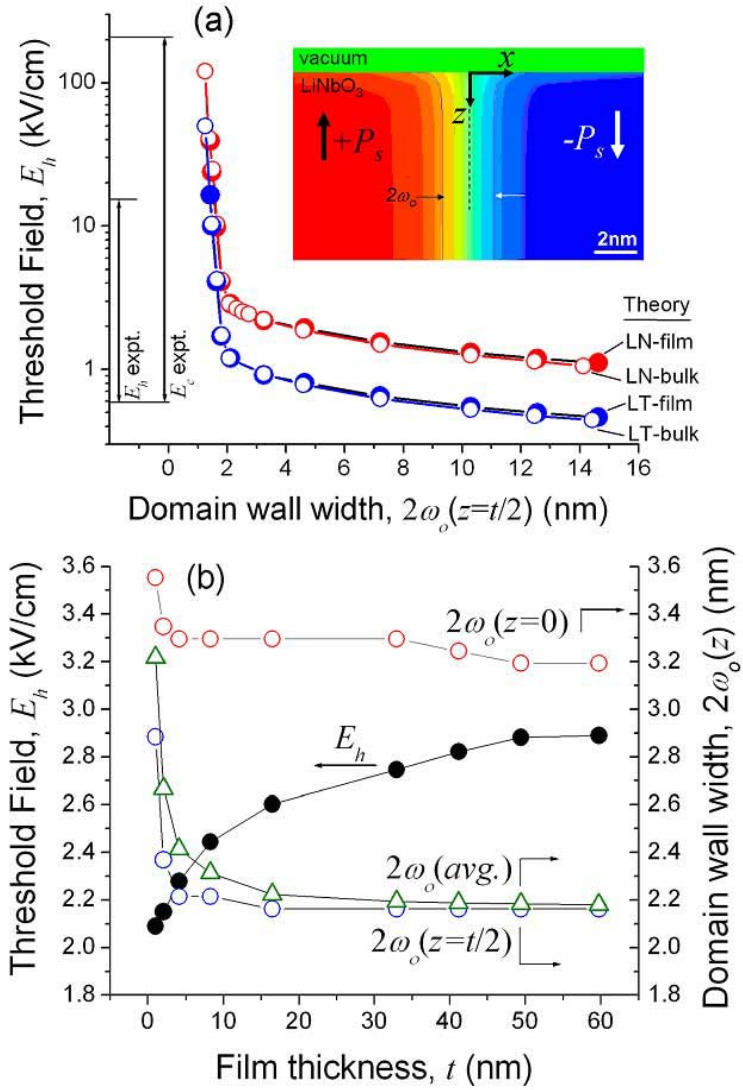

**Figure 1(a):** Phase field modeling of the threshold coercive field, $E_h$ for the motion of a single domain wall in LiNbO$_3$ (LN) and LiTaO$_3$ (LT) versus domain wall width, $2\omega_o$. The simulation size was 128$a$x2$a$x128$a$ for $2\omega_o$ <2nm and 512$a$x2$a$x128$a$ for $2\omega_o$ >2nm, where $a$=0.515nm. The film was $t$=96$a$ thick. Inset shows the polarization profile at the junction between the wall and one of the surfaces of the film. The bulk simulation was infinite in all dimensions. **(b):** The dependence of $E_h$ on film thickness for a fixed $g_{13}$=2$\alpha_{33}a^2$. Also the domain wall, 2 $\omega_o$ at $z$=0, $z$=$t/2$ and an average quantity, $2\int \omega_o(z)dz/t$ are plotted.

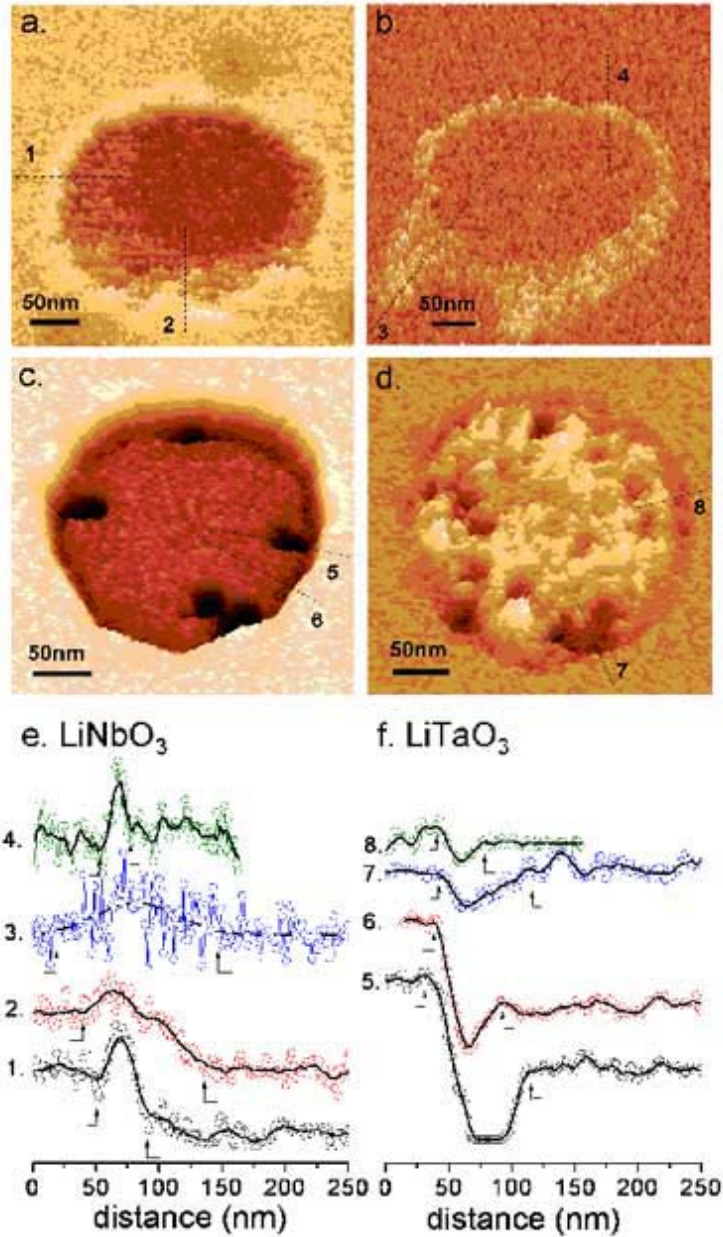

**Figure 2:** Scanning nonlinear dielectric microscopy images of a circular domain in a 40nm thick z-cut single crystal lithium niobate (a,b) and a 31nm thick z-cut single crystal lithium tantalate, (c,d) at first harmonic, $\omega_p$ =6kHz (a,c), and second harmonic, $2\omega_p$ =12kHz (b,d) modulation frequencies. Figure 2d, e show typical line profiles from these images as labeled, and pairs of arrows indicate the wall region. Images (a-d) are plotted in 3-dimensional orthographic view with a ~21° rotation about the horizontal axes of the images. The length scales directly correspond to the horizontal axes of the images. The same domain region is imaged in (a) and (b), and similarly in (c) and (d).